\documentclass[journal]{IEEEtran}

\usepackage{cite}
\usepackage{url}
\usepackage{graphicx}
\usepackage{color}
\usepackage{float}
\usepackage{tabularx,colortbl}
\usepackage{ifthen}

\usepackage{amsmath}
\usepackage{amsthm,amsfonts}
\usepackage{pgfplots}
\usepackage[normalem]{ulem}

\usepackage{tikz}
\usepackage[customcolors]{hf-tikz}
\usepgfplotslibrary{colorbrewer}
\usetikzlibrary{mindmap,trees}
\usetikzlibrary{%
  3d,
  arrows,%
  shapes.misc,
  shapes.arrows,%
  chains,%
  matrix,%
	fit,%
  positioning,
  scopes,%
  decorations.pathmorphing,
  shadows,%
  decorations.markings,
  shapes.geometric,
  arrows.meta,bending,
	patterns,
	intersections, 
	pgfplots.fillbetween
}

\usepackage{tikz-3dplot}
\usepackage[pdfusetitle]{hyperref}
\hypersetup{
    colorlinks,
    linkcolor={red!50!black}, 
    citecolor={blue!50!black},
    urlcolor={blue!50!black}
}

\usepackage{cellspace}
\newcommand\cincludegraphics[2][]{\raisebox{-0.3\height}{\includegraphics[#1]{#2}}}

\providecommand*{\mat}[1]{\mathbf#1}
\providecommand*{\M}[1]{\mathbf#1}
\providecommand*{\tM}[1]{\tilde{\mathbf#1}}
\providecommand*{\mrm}[1]{\mathrm{#1}}
\providecommand*{\unit}[1]{\ensuremath{\mrm{\,#1}}}
\renewcommand{\vec}[1]{{\boldsymbol#1}}
\providecommand*{\V}[1]{\boldsymbol#1}
\providecommand*{\UV}[1]{\hat{\boldsymbol#1}}
\providecommand*{\T}[1]{\mathrm{#1}}
\DeclareMathAccent{\ring}{\mathalpha}{operators}{"17}

\providecommand*{\unit}[1]{\ensuremath{\mrm{\,#1}}}

\providecommand*{\ju}{\ensuremath{\mrm{j}}}

\renewcommand{\Re}{\operatorname{Re}}	
\providecommand*{\diff}{\operatorname{d}\!}

\providecommand*{\diffV}{\operatorname{dV}\!}

\newcommand{\Tr}{\mathop{\mrm{Tr}}\nolimits}

\newcommand{\ave}[1]{\langle#1\rangle} 
\newcommand{\norm}[1]{\lVert#1\rVert} 

\newcommand{\R}{\mathbb{R}{}}

\newcommand{\ie}{\textit{i.e.}\/, }
\newcommand{\eg}{\textit{e.g.}\/, }
\newcommand{\cf}{\textit{cf.}\/, }

\newcommand{\trans}{\text{T}}
\newcommand{\herm}{\text{H}}
\newcommand{\Id}{\mat{1}}

\newcommand{\maximize}{\mrm{maximize}}

\newcommand{\subto}{\mrm{subject\ to}}

\newcommand{\rhor}{\rho_\mrm{r}}

\newcommand{\radm}{\varrho} 

\newcommand{\Aeff}{A_\mrm{eff}}
\newcommand{\As}{A_\mrm{s}}

\newcommand{\Na}{N_\mrm{A}} 
\newcommand{\Nr}{N_1} 
\newcommand{\reg}{\varOmega}
\newcommand{\breg}{\partial\reg}
\newcommand{\rv}{\vec{r}}

\colorlet{dpurple}{blue!50!red}
\colorlet{dblue}{blue!50!black}
\colorlet{dgreen}{green!50!black}
\colorlet{dred}{red!50!black}
\colorlet{dyellow}{yellow!50!black}
\colorlet{dorange}{orange!50!black}
\definecolor{metal}{RGB}{218,165,32}
\definecolor{diel}{RGB}{1,165,32}
\definecolor{antenna}{RGB}{100,150,162}
\definecolor{breg}{rgb}{0.2,0.6,0.8}%
\definecolor{preg}{rgb}{0.8,0.2,0.2}%
\definecolor{reg}{RGB}{218,165,32}

\graphicspath{{figures/}{tikz/}}

\makeatother

\title{Degrees of freedom for radiating systems}

\author{
Mats Gustafsson
\thanks{Manuscript received \today; revised \today. This work was supported by ELLIIT - an Excellence Center at Linkoping-Lund in Information Technology and the TICRA foundation.}
\thanks{M. Gustafsson is with Lund University, Lund, Sweden, (e-mails: mats.gustafsson@eit.lth.se).}
}

\tdplotsetmaincoords{70}{140}
\pgfmathsetmacro{\wx}{0.8} 
\pgfmathsetmacro{\wy}{3.1} 
\pgfmathsetmacro{\wz}{0.4} 
\pgfmathsetmacro{\fx}{0} 
\pgfmathsetmacro{\fy}{1}
\pgfmathsetmacro{\fz}{0.4}
\pgfmathsetmacro{\fw}{0.2}
\pgfmathsetmacro{\hx}{2} 
\pgfmathsetmacro{\hy}{2.7}
\pgfmathsetmacro{\hz}{1}
\pgfmathsetmacro{\d}{0.2}
\tikzset{%
		grid/.style={very thin,gray},
		axis/.style={->,white,thin},
		cube/.style={fill=metal,fill opacity=0.5,draw=blue!50!black},
    horn/.style={fill=black!50!blue,fill opacity=0.5,draw=blue!50!black},
		cube hidden/.style={fill=black!50!blue,fill opacity=0.5,draw=blue!50!black},
    xyplane/.style={canvas is xy plane at z=#1,very thin}    
    }

\begin{document}

\maketitle
\begin{abstract}
Electromagnetic degrees of freedom are instrumental in antenna design, wireless communications, imaging, and scattering. Larger number of degrees of freedom enhances control in antenna design, influencing radiation patterns and directivity, while in communication systems, it links to spatial channels for increased data rates and reliability, and resolution in imaging. The correlation between computed degrees of freedom and physical quantities is not fully understood, prompting a comparison between classical estimates, Weyl's law, modal expansions, and optimization techniques. In this paper, it is shown that the number of degrees of freedom for arbitrary shaped radiating structures approaches the shadow area measured in squared wavelengths asymptotically as the wavelength decreases.
\end{abstract}

\section{Introduction}
\IEEEPARstart{E}{lectromagnetic} degrees of freedom (DoF) finds importance across a wide range of electromagnetic (EM) applications, spanning antenna design, wireless communication systems, scattering problems, and measurement techniques~\cite{Bucci+Franceschetti1989,Bucci+Isernia1997,Piestun+Miller2000,Migliore2006a,Migliore2008,Kildal+etal2017,Hu+etal2018,Ehrenborg+Gustafsson2020,Pizzo+etal2022,Gustafsson+Lundgren2024,Miller2019}. The number of DoF (NDoF) emerges as a crucial factor in achieving desired performance levels. For instance, in antenna design, a higher NDoF facilitates enhanced control over radiation patterns and maximum directivity~\cite{Kildal+etal2017}. In wireless communication systems, NDoF correlates with the number of channels or spatial dimensions utilized for signal transmission, thereby providing higher data rates and improved reliability, especially in technologies like multiple-input multiple-output (MIMO) systems and intelligent surfaces~\cite{Pizzo+etal2022,Hu+etal2018}. 

The relationship between the NDoF and the electrical size of an object have been explored for at least half-a-century investigating resolution in imaging systems~\cite{DiFrancia1955} and antenna systems~\cite{DiFrancia1956}. Small electrical sizes exhibit few DoF while large sizes display numerous DoF and hence, \eg potentially higher directivity and narrow beamwidth. 
Analytically solvable cases such as spherical regions with radius $a$ offer a simple estimate of $2(ka)^2$ DoFs~\cite{Bucci+Isernia1997} based on the first $L\approx ka$ spherical modes for wavenumber $k=2\pi/\lambda$, wavelength $\lambda$, and two polarization DoFs. The number of propagating waves in hollow rectangular waveguides~\cite{Hill1994} are also solvable with asymptotically $2\pi A/\lambda^2$ DoFs for a cross-sectional area $A$. These results are traced back to Weyl more than a century ago, who studied the eigenvalue distribution of the Laplace and Helmholtz operators~\cite{Weyl1911,Arendt+etal2009}. Alternative estimates based on Nyquist ($\lambda/2$) sampling resulting in $8A/\lambda^2$ DoFs have also been extensively used~\cite{Bucci+Isernia1997,Bucci+etal1998,Migliore2006a,Migliore2008,Franceschetti2017}. Moreover, communication systems have recently been investigated for different configurations~\cite{Pizzo+etal2022,Hu+etal2018}, with $2\pi A/\lambda^2$ DoFs for planar apertures.

In this paper, Weyl's law~\cite{Weyl1911,Arendt+etal2009} and its connection to the NDoF are first explored. Beyond its mathematical significance in spectral theory, Weyl's law is also fundamental in explaining the NDoF in thermal radiation~\cite{Arendt+etal2009}. However, its relevance to the NDoF in antenna and communication systems has been largely overlooked. Weyl's law demonstrates that the NDoF for arbitrarily shaped bodies scale with the surface area, measured in squared wavelengths. These NDoF can be interpreted as the DoF in a communication system between the body and the surrounding region. 
Weyl's law provides a unified framework that extends previous results on the asymptotic NDoF of canonical shapes, such as spheres~\cite{Bucci+Isernia1997} and planar rectangles~\cite{Pizzo+etal2022,Hu+etal2018}, to arbitrarily shaped regions.

In antenna systems, it is more common to focus on radiating systems, where the antenna transmits energy to the far-field. This paper shows that the radiating NDoF approaches the total shadow area of the region, measured in squared wavelengths. For convex shapes, this result matches the NDoF derived from Weyl's law, but it differs for more general shapes. 
The reduction in NDoF for non-convex shapes can be attributed to the blockage of some regions, which limits their contribution to the radiated fields. 

The NDoF for radiating systems is crucial in antenna and communication system design, as it determines the maximum number of radiating fields that can be achieved within a constrained design region. Furthermore, it is directly related to the maximum achievable average gain of an antenna system when its beam is steered across all possible directions. This establishes a connection to the Hannan limit, which was originally derived for infinite planar arrays~\cite{Hannan1964,Kildal+etal2015}. 
Notably, this radiating NDoF is also twice the number of significant characteristic modes~\cite{Gustafsson+Lundgren2024}. In communication systems, the derived asymptotic limit offers a straightforward estimate of the NDoF, and thus the capacity~\cite{Franceschetti2017}. This generalizes the results previously obtained for planar rectangular regions~\cite{Pizzo+etal2022,Hu+etal2018} to arbitrarily shaped regions. The possibility of extending this framework to communication between arbitrarily shaped regions has also been explored~\cite{Gustafsson2024c}.

The paper is organized as follows. Section~\ref{S:Weyl} introduces Weyl's law and its applications for determining NDoF. Section~\ref{S:capacity} explores the concept of capacity and its connection to radiation modes and NDoF. In Section~\ref{S:NDoFshadowarea}, the asymptotic NDoF is analyzed in relation to the shadow area of a region. Electric and magnetic currents are examined in Section~\ref{S:EMcurrents}. NDoF in inverse problems is discussed in Section~\ref{S:InverseProblems}. 
Finally, the paper is concluded in Section~\ref{S:conclusion}.

\section{Weyl's law and propagating modes}\label{S:Weyl}
NDoF is most straight forward determined for the number of propagating modes in a waveguide. Canonical shapes such as rectangular or circular cross sections can be solved analytically~\cite{Hill1994}. These and arbitrary shaped cross sections can also be treated as a special case of Weyl's law~\cite{Weyl1911,Arendt+etal2009}.
Weyl's law describes the distributions of eigenvalues $\nu_n$ for the Laplace operator $-\nabla^2 u_n=\nu_n u_n$ in a region $\reg\subset\R^{d}$ with Dirichlet or Neumann boundary conditions. The number of eigenvalues $N_{\T{W}d}(\nu)$ below $\nu$ is asymptotically~\cite{Weyl1911,Arendt+etal2009} 
\begin{equation}
	N_{\T{W}d}(\nu) =  \frac{w_d |\varOmega|\nu^{d/2}}{(2\pi)^{d}}
	\quad\text{as } \nu\to\infty
\label{eq:WeylLap}
\end{equation}
for a region $\reg$ with volume $|\reg|$ (in $\R^3$ or area in $\R^2$ or length in $\R^1$) and with $w_d=\pi^{d/2}/(d/2)!$ denoting the volume of the unit ball in $\R^{d}$. Weyl's law is also applicable to the number of positive eigenvalues for Helmholtz equation in $\reg$
\begin{equation}
	\nabla^2 u_n + k^2 u_n= \mu_n u_n
	\Leftrightarrow
	-\nabla^2 u_n = (k^2-\mu_n)u_n
\label{eq:WeylHelmholtz}
\end{equation}
for a wavenumber $k=2\pi/\lambda$ with wavelength $\lambda$.
The number of propagating $k^2>\mu$ (scalar) modes in arbitrary shaped waveguides is hence asymptotically 
\begin{equation}
	N_{\T{W}d}(k^2) = \frac{w_d |\varOmega| k^{d}}{(2\pi)^{d}}
	= \frac{w_d |\varOmega|}{\lambda^d}
\label{eq:WeylPropModes}
\end{equation}
which reduces to
\begin{equation}
	N_{\T{W}1} = \frac{k\ell}{\pi}=\frac{2\ell}{\lambda}
	\quad\text{and }
	N_{\T{W}2} = \frac{k^2 A}{4\pi} = \frac{\pi A}{\lambda^2}
\label{eq:Weyl}
\end{equation}
for a line with length $\ell=|\reg|$ in $\R$ and an aperture surface with area $A=|\reg|$ in $\R^2$. Here, we note that the 1D case corresponds to Nyquist ($\lambda/2$) sampling. The 2D case differs from Nyquist sampling in two orthogonal directions, even for the special case of a rectangular region~\cite{Hill1994,Hu+etal2018}. Explicit evaluation of a rectangular cross section provides a simple interpretation of the difference between $\lambda/2$ for lines and the area scaling~\eqref{eq:Weyl} for surfaces. Waves in 1D start to propagate for sizes above half-a-wavelength. For a rectangular waveguide propagating modes depend on the squared magnitude of oscillations in two transverse directions corresponding to the area of an ellipse or circle~\cite{Hill1994}. This difference is quantified by the unit ball $w_d$ in Weyl's law~\eqref{eq:WeylPropModes}, \ie $w_1^2=4$ and $w_2=\pi$.   

These estimates~\eqref{eq:Weyl} of the asymptotic number of propagating modes for (scalar) Helmholtz equation are doubled for Maxwell's equations when two orthogonal polarizations contribute
\begin{equation}
	\Na = \frac{k^2 A}{2\pi} = \frac{2\pi A}{\lambda^2}
\label{eq:Weyl2D}
\end{equation}
for an aperture with area $A\in\R^2$. The NDoF~\eqref{eq:Weyl2D} is valid for arbitrary shaped cross sections in the electrically large limit $k\to\infty$. 
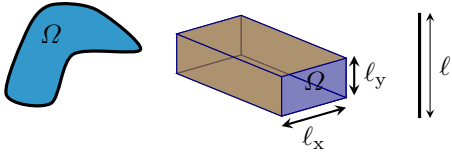
\begin{figure}%
{\centering
\begin{tikzpicture}[scale=0.7,>= stealth]
	\begin{scope}[xshift=-3cm,yshift=-1mm]
	\draw [fill=breg,very thick] plot [smooth cycle,scale=0.85] coordinates {(-1,-1) (0,-1.2) (0.5,-0.1) (2,0) (1,0.9) (0,1) (-0.5,0.5)};    
	\node at (0,0.3) {$\reg$};	
	\end{scope}
\begin{scope}[tdplot_main_coords]
  \draw[cube hidden] (-\wx,0,-\wz) -- (\wx,0,-\wz) -- (\wx,0,\wz) -- (-\wx,0,\wz) -- cycle; 
  \draw[cube hidden] (-\wx,0,-\wz) -- (-\wx,\wy,-\wz) -- (-\wx,\wy,\wz) -- (-\wx,0,\wz) -- cycle; 
	\draw[cube hidden] (-\wx,0,-\wz) -- (-\wx,\wy,-\wz) -- (\wx,\wy,-\wz) -- (\wx,0,-\wz) -- cycle; 
	\draw[cube] (\wx,0,-\wz) -- (\wx,\wy,-\wz) -- (\wx,\wy,\wz) -- (\wx,0,\wz) -- cycle; 
	\draw[cube,fill=none] (-\wx,\wy,-\wz) -- (\wx,\wy,-\wz) -- (\wx,\wy,\wz) -- (-\wx,\wy,\wz) -- cycle; 
	\draw[cube] (-\wx,0,\wz) -- (-\wx,\wy,\wz) -- (\wx,\wy,\wz) -- (\wx,0,\wz) -- cycle;

\begin{scope}[shift={(0,\wy,0)}]  
\draw[<->,thick] (-\wx,0,-\wz-\d) -- node[anchor=north] {$\ell_{\mrm{x}}$} (\wx,0,-\wz-\d);
\draw[<->,thick] (-\wx-\d,0,-\wz) -- node[anchor=west] {$\ell_{\mrm{y}}$} (-\wx-\d,0,\wz);
\end{scope}
\end{scope}
\begin{scope}[xshift=4cm,yshift=-4mm]
	\draw[very thick] (0,-1) -- +(0,2);
	\draw[<->] (0.2,-1) -- node[right] {$\ell$} +(0,2);
\end{scope}	
	\node at (2,-0.7) {$\reg$};	
\end{tikzpicture}
\vspace{-4mm}
\par}
\caption{The number of eigenvalues for Laplace and Helmholtz operators in a region $\reg$ is described by Weyl's law as applied to waveguides and line elements in~\eqref{eq:Weyl}.}%
\label{fig:Weylslaw}%
\end{figure}

Weyl's law~\eqref{eq:Weyl2D} estimates the number of propagating modes in electromagnetic waveguide structures with perfect electric conductor (PEC) walls, but it is more challenging to apply Weyl's law for radiating structures in $\R^3$. Starting with a spherical region with radius $a$ which is analytically solvable by expanding the radiated field in spherical waves~\cite{Harrington1961,Kristensson2016}. In contrast to waveguides, there is no clear cut-off between propagating and evanescent waves. There are $2L(L+2)$ spherical modes up to order $L$, and modes of order $L\geq ka$ become increasingly reactive~\cite{Harrington1961}. Using the surface area of the sphere together with $L\approx ka$ corresponds to approximately~\cite{Bucci+Isernia1997}
\begin{equation}
 2(ka)^2=\frac{k^2 A}{2\pi}=\frac{2\pi A}{\lambda^2}
\label{eq:NDoFsph}
\end{equation}
propagating modes for $ka\gg 1$ similar to the estimate from Weyl's law~\eqref{eq:Weyl2D} for an aperture in $\R^2$. 

An explanation for the absence of a clear transition between propagating and evanescent modes in the context of a sphere arises from its expanding surface area as the radial distance increases. Consequently, there is a proportional rise in the number of propagating modes with radial distance. This stands in sharp contrast to waveguiding structures, where the cross-sectional area, and thus the NDoF, remains constant along the propagation direction. Hence, the approximation $L\approx ka$ emerges as a useful descriptor for the shift from predominantly propagating to evanescent waves.

Although the cutoff $L\approx ka$ provides a good simple approximation for the order of radiating spherical waves. Improved estimates for near-field measurements~\cite{Hansen1988} and computational techniques~\cite{Song+Chew2001b} can be used for more precise estimates. An alternative interpretation of the cutoff is that high-order modes are associated with large current amplitudes and hence low efficiency for radiators of finite conductivity. 
This is related to the rapidly increasing reactive near field for high order modes, \ie there are large field (and current) amplitudes that do not contribute efficiently to the radiated power. These fields are associated with high losses and stored energy for antennas. This imbalance between the amplitude of the modes can also be interpreted as a communication channel with large dynamical range between the singular values of the channel matrix requiring a high SNR.  

\begin{figure}[t]%
\includegraphics[width=\columnwidth]{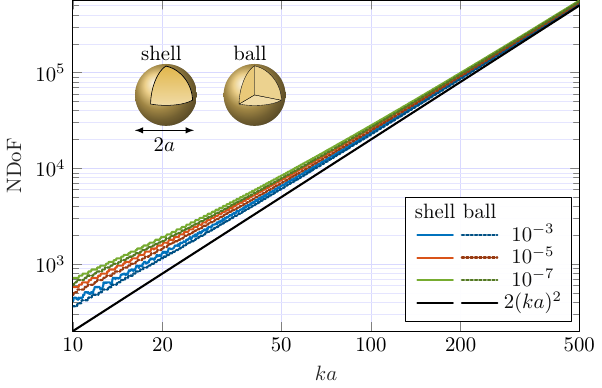}%
\caption{NDoF from spherical modes with at least $50\%$ efficiency for spherical shells with surface resistivity $R_\mrm{s}=10^{-p}\eta_0$ and balls with resistivity $\rhor=10^{-p}\eta_0/k$ and both having radius $a$ and $p=3,5,7$. The NDoF is compared with the estimate $2(ka)^2$ from~\eqref{eq:NDoFsph}.}%
\label{fig:NDoFsph_tols_ka}%
\end{figure}

Evaluating the number of radiating spherical waves with efficiency above a given threshold, here $50\%$, produces an estimate of the NDoF. Fig.~\ref{fig:NDoFsph_tols_ka} depicts the number of such radiating modes~\cite{Ehrenborg+Gustafsson2020} on a spherical shell with surface resistivity $R_{\mrm{s}}=10^{-p}\eta_0$ and spherical ball with resistivity $k\rho_{\mrm{r}}=10^{-p}\eta_0$ for $p=3,5,7$. Note that copper has a surface resistivity of approximately $R_{\mrm{s}}\approx 0.01\unit{\Omega/\square}$ around $1\unit{GHz}$ and a resistivity of $\rhor\approx 10^{-8}\unit{\Omega\, m}$. The results are compared with the estimate~\eqref{eq:NDoFsph}, and it is observed that the relative difference between the computed NDoF and $2(ka)^2$ decreases as $ka\to\infty$.
The radiation efficiencies for the spherical regions used in Fig.~\ref{fig:NDoFsph_tols_ka} are determined by expanding the current density and fields in spherical waves. The solution is expressed in radiation modes~\cite{Schab2016,Gustafsson+etal2020} as given in App.~\ref{S:RadiationModes}.
Using material losses opens a possibility to generalize the NDoF to arbitrary shaped radiators as analyzed in this paper.

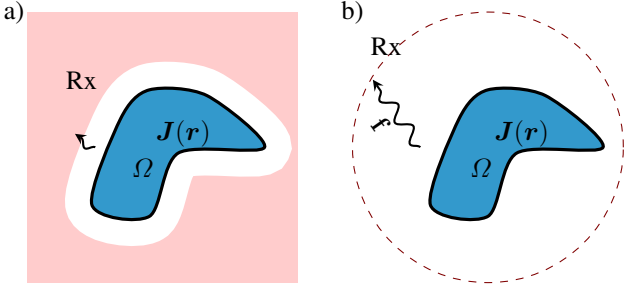
\begin{figure}[t]%
{\centering
\begin{tikzpicture}[scale=0.9,>= stealth]
	\begin{scope}
	\fill[red!20!white] (-1.8,-2) rectangle (2.2,2);
	\draw [fill=white,white,line width=20pt] plot [smooth cycle,scale=0.85] coordinates {(-1,-1) (0,-1.2) (0.5,-0.1) (2,0) (1,0.9) (0,1) (-0.5,0.5)};    
	\draw [fill=breg,very thick] plot [smooth cycle,scale=0.85] coordinates {(-1,-1) (0,-1.2) (0.5,-0.1) (2,0) (1,0.9) (0,1) (-0.5,0.5)};    
	\node at (-0.1,-0.3) {$\reg$};	
	\node at (0.5,0.2) {$\vec{J}(\rv)$};	
\draw[->,thick,decorate,decoration=snake] (-0.8,0) -- (-1.1,0.2);	
	\node at (-1,1) {Rx};	
	\node at (-2,2) {a)};
	\end{scope}	
	\begin{scope}[xshift=5cm]
	\draw [fill=breg,very thick] plot [smooth cycle,scale=0.85] coordinates {(-1,-1) (0,-1.2) (0.5,-0.1) (2,0) (1,0.9) (0,1) (-0.5,0.5)};    
	\node at (-0.1,-0.3) {$\reg$};	
	\node at (0.5,0.2) {$\vec{J}(\rv)$};	
\draw[->,thick,decorate,decoration=snake] (-1,0) -- node[sloped,below] {$\M{f}$} (-1.7,1);	
\draw[dashed,dred] (0,0) circle(2cm);
	\node at (-1.5,1.5) {Rx};	
	\node at (-2,2) {b)};	
	\end{scope}	
\end{tikzpicture}
\vspace{-1mm}
\par}
\caption{DoF for a region $\reg\in\R^3$. (left) radiation to a volume surrounding $\reg$ related to Weyl's law~\eqref{eq:Weyl2D}. (right) radiation to the far field or a circumscribing sphere.}%
\label{fig:RDoF}%
\end{figure}

Arbitrarily shaped closed surfaces and disjoint regions are much more complex, for which it is also apparent that there are several alternative ways to interpret DoF, such as communication between regions~\cite{Piestun+Miller2000,Jensen+Wallace2008,Miller2019,Gustafsson2024c}, from a region to a volume surrounding the region in Fig.~\ref{fig:RDoF}a, and from a region to the far field in Fig.~\ref{fig:RDoF}b. 
In this paper, we consider radiating NDoF interpreted as the NDoF from a region to the far field.

For arbitrary shaped regions with smooth boundary, we can interpret Weyl's law locally, resulting in a NDoF according to~\eqref{eq:Weyl2D} with surface area $A$. However, these DoF will generally only propagate to the volume outside $\reg$ and not to the far field. Instead, the NDoF can be considered to propagate from the region $\reg$ to a slightly enlarged version of $\reg$ as illustrated in Fig.~\ref{fig:RDoF}a. This short distance is considered fixed and translates to infinity many wavelengths in the high-frequency limit. Also note that the NDoF is unbounded if this distance is removed.

\section{Capacity limits and radiation modes}\label{S:capacity}
Treating arbitrary shaped (non-convex or non-connected) regions $\reg\in\R^3$, we can formulate a communication problem between transmitters in the region $\reg$ to receivers in the far field, or equivalently, on a circumscribing sphere, as shown in Fig.~\ref{fig:RDoF}b. Current density $\V{J}(\rv)$ in the transmitting region $\reg$ constitutes sources for the radiated field. These currents can be expanded in sufficiently large number of basis functions, with expansion coefficients collected in a column matrix $\M{I}$~\cite{Harrington1968}. Similarly, the radiated field is expanded in spherical waves~\cite{Harrington1961,Kristensson2016}, with expansion coefficients collected in a column matrix $\M{f}=-\M{U}\M{I}$, where $\M{U}$ denotes the projection of spherical waves onto the used basis functions~\cite{Gustafsson+Nordebo2013}. The capacity (spectral efficiency) of this idealized system $\M{f}=-\M{U}\M{I}+\M{n}$ is determined by~\cite{Ehrenborg+Gustafsson2020}
\begin{equation}
\begin{aligned}
& \maximize &&  \log_2\left(\det\left(\Id+\gamma\M{U}\M{P}\M{U}^{\herm}\right)\right) \\
& \subto && \Tr(\M{R}\M{P})=1 \\
& && \M{P} \succeq \M{0},
\end{aligned}
\label{eq:MaxCapI1}
\end{equation}
where $\M{P}=\mathcal{E}\{\M{I}\M{I}^{\herm}\}$ denotes the covariance matrix of the currents, $\M{R}$ is the resistance matrix~\cite{Harrington1968}, and $\gamma$ quantifies the signal-to-noise ratio (SNR) from the additive noise $\M{n}$. In~\eqref{eq:MaxCapI1}, the current is normalized to (two times) unit dissipated power~\cite{Ehrenborg+Gustafsson2020}. To provide more realistic values for the capacity, physical properties can be considered. For antennas, it is common to incorporate limitations originating in the efficiency~\cite{Ehrenborg+Gustafsson2018,Ehrenborg+Gustafsson2020} or bandwidth from the reactive energy around the antenna~\cite{Ehrenborg+Gustafsson2018,Ehrenborg+etal2021}. 

The maximum capacity~\eqref{eq:MaxCapI1} is determined by water filling~\cite{Paulraj+etal2003} for which it is convenient to rewrite~\eqref{eq:MaxCapI1} in a standard form by a change of variable $\tM{P}=\M{G}\M{P}\M{G}^{\herm}$, where $\M{R}=\M{G}^{\herm}\M{G}$ is a factorization of the resistance matrix. Substituting into~\eqref{eq:MaxCapI1} 
\begin{equation}
\begin{aligned}
& \maximize &&  \log_2\left(\det\left(\Id+\gamma\M{H}\tM{P}\tM{H}^{\herm}\right)\right) &&&\\
& \subto && \Tr(\tM{P})=1 &&&\\
& && \tM{P} \succeq \M{0}&&&
\end{aligned}
\label{eq:MaxCapI2}
\end{equation}
with the channel matrix $\M{H}=\M{U}\M{G}^{-1}$. This problem is diagonalized by a singular value decomposition (SVD) of the channel matrix $\M{H}$, \ie (square roots of the) eigenvalues of $\M{H}^{\herm}\M{H} = \M{G}^{-\herm}\M{R}_0\M{G}^{-1}$, which can be written $\M{G}^{-\herm}\M{R}_0\M{G}^{-1}\M{V}_n = \nu_n\M{V}_n$, where $\M{U}^{\herm}\M{U}=\M{R}_0$ is identified as the radiation matrix~\cite{Gustafsson+etal2020}. Simplifying by multiplication with $\M{G}^{\herm}$ and setting $\M{I}_n=\M{G}^{-1}\M{V}_n$ results in an eigenvalue problem for radiation mode efficiency
\begin{equation}
	\M{R}_0\M{I}_n = \nu_n\M{R}\M{I}_n = \nu_n(\M{R}_0+\M{R}_{\rho})\M{I}_n.
\label{eq:RadmEff}
\end{equation}
The resistance matrix $\M{R}=\M{R}_0+\M{R}_{\rho}$ is here decomposed into the radiation matrix $\M{R}_0=\M{U}^{\herm}\M{U}$ modelling radiated power $P_{\mrm{r}}=\frac{1}{2}\M{I}^{\herm}\M{R}_0\M{I}$ and the material matrix $\M{R}_\rho$ modelling Ohmic or dielectric losses $P_{\mrm{\rho}}=\frac{1}{2}\M{I}^{\herm}\M{R}_\rho\M{I}$.

\begin{figure}%
\includegraphics[width=\columnwidth]{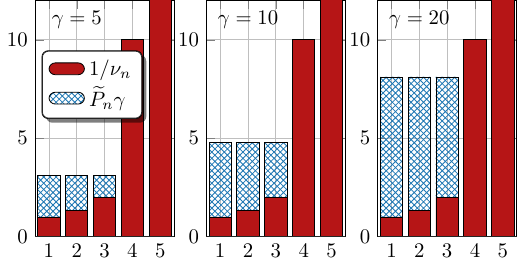}%
\caption{Illustration of the water filling solution~\eqref{eq:CapRadm} for radiation mode efficiencies $\nu_n\in\{1, 0.75, 0.5, 0.1, 0.01\}$. The red bars represent $1/\nu_n$, while the hashed blue bars indicate the allocated power $\tilde{P}_n\gamma$ for the different modes. }%
\label{fig:waterfilling}%
\end{figure}

Radiating modes diagonalizes~\eqref{eq:MaxCapI2}, which solved using waterfilling over radiation mode efficiencies $\nu_n$ results in
\begin{equation}
	\max_{\sum \tilde{P}_n=1}\sum_{n=1}^{\infty} \log_2\left(1+\gamma \nu_n\tilde{P}_n\right),
\label{eq:CapRadm}
\end{equation}  
with a finite number of non-zero power levels $\tilde{P}_n\geq 0$ associated with radiation modes of sufficiently high efficiency. The number of used modes depends on the signal-to-noise ratio $\gamma$ and the distribution of the modes $\nu_n$. Waterfilling solutions for five efficiencies $\nu_n\in\{1, 0.75, 0.5, 0.1, 0.01\}$ and SNRs $\gamma\in\{5,10,20\}$ are depicted in Fig.~\ref{fig:waterfilling}. 
The reciprocals of the efficiencies, $1/\nu_n$, are large for inefficient modes, \eg mode 4 has $1/\nu_4=10$, and mode 5 has $1/\nu_5=100$. The normalized power level allocated to each mode, $\tilde{P}_n \gamma$, in the water-filling algorithm can be interpreted as determining the water level over a bottom profile defined by $1/\nu_n$, with a total amount of water $\gamma = \sum \tilde{P}_n \gamma$. The allocation of normalized power $\tilde{P}_n \gamma$ and the number of active modes depend on the SNR, $\gamma$. 
For low SNR, $\gamma \ll 1$, only the strongest mode, with $1/\nu_1=1$, contributes. As $\gamma$ increases, power is quickly allocated to additional modes, such as modes 2 and 3, with $1/\nu_2 \approx 1.33$ and $1/\nu_3 = 2$, respectively. Inefficient modes, such as $1/\nu_4=10$, require significantly higher SNR to be activated, as shown in Fig.~\ref{fig:waterfilling}.

The capacity~\eqref{eq:CapRadm} is solely determined by the efficiencies $\nu_n$ of the radiation modes and the signal-to-noise ratio (SNR) $\gamma$. Consequently, radiation modes can be regarded as fundamental physical quantities that quantify radiation properties and degrees of freedom for arbitrary-shaped objects with material losses~\cite{Ehrenborg+Gustafsson2020}.

Geometry and material properties are encapsulated within the radiation modes~\eqref{eq:RadmEff}. To facilitate analysis, it is advantageous to distinguish between the radiation contribution $\M{R}_0$ and the material contribution $\M{R}_\rho$. Radiation mode eigenvalues are defined by the generalized eigenvalue problem~\cite{Schab2016}
\begin{equation}
\M{R}_0\M{I}_n = \varrho_n\M{R}_{\rho}\M{I}_n, 
\label{eq:radmeig}
\end{equation}
with eigenvalues ordered decreasingly, \ie $\radm_{n}\geq\radm_{n+1}$, and efficiencies $\nu_n=\radm_n/(1+\radm_n)$ in~\eqref{eq:RadmEff}. These modal currents are orthogonal over the material loss matrix $\M{I}_m^{\herm}\M{R}_{\rho}\M{I}_n=\delta_{mn}$ and radiated far fields $\M{f}_m^{\herm}\M{f}_n=\M{I}_m^{\herm}\M{R}_0\M{I}_n=\radm_n\delta_{mn}$. The number of radiation modes are infinite, but they diminish rapidly as $\radm_n\to 0$ for large $n$, similarly to the spherical waves in~\eqref{eq:NDoFsph}, to which they also reduce for spherical regions~\cite{Gustafsson+etal2020}. Radiation modes~\eqref{eq:radmeig} are orthogonal and maximize the (Rayleigh) quotient between the radiated power and dissipated power in materials
\begin{equation}
	\frac{\mrm{radiated\ power}}{\mrm{dissipated\ power\ in\ material}}
	=
	\frac{\M{I}_m^{\herm}\M{R}_0\M{I}_n}{\M{I}_m^{\herm}\M{R}_{\rho}\M{I}_n}
	=\delta_{mn}\radm_n
\label{eq:Rayleighradmod}
\end{equation}
which can be used to define NDoF from sufficiently efficient modes. Setting a threshold of $50\%$ efficiency corresponding to radiation modes $\radm_n\geq 1$, which is here used to define a NDoF $\Nr$ 
for arbitrary shaped regions and material losses~\cite{Ehrenborg+Gustafsson2020}. 

\begin{figure}%
\includegraphics[width=\columnwidth]{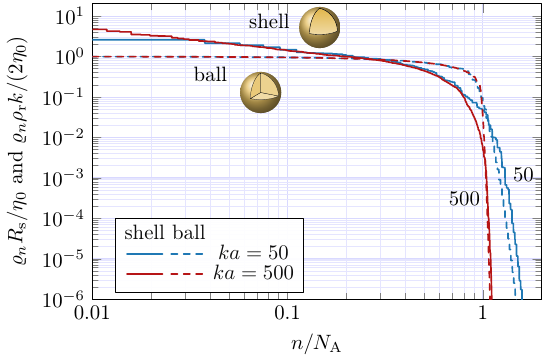}%
\caption{Normalized radiation modes for spherical shells and spherical balls of the electrical sizes $ka=50$ and $ka=500$. The mode index is normalized by the asymptotic NDoF $\Na=2(ka)^2$.}%
\label{fig:radmodessph}%
\end{figure}

\begin{figure}%
\includegraphics[width=\columnwidth]{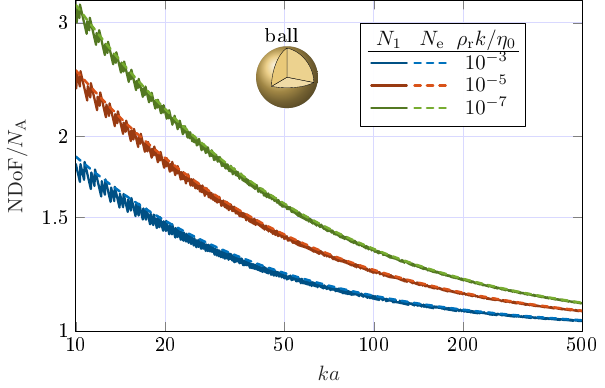}%
\caption{Comparison between the NDoF $\Nr$ defined by $50\%$ efficiency ($\radm_n\geq 1$) and the effective NDoF $N_\T{e}$ in~\eqref{eq:eNDoF}. The NDoFs are normalized by $\Na=2(ka)^2$ for spherical balls with resistivity $\rho_\mrm{r}=10^{-n}\eta_0/k$ with $n=3,5,7$.}%
\label{fig:NDoFsphN_tols_ka}%
\end{figure}

Fig.~\ref{fig:radmodessph} depicts normalized radiation modes for spherical shells and spherical balls of the electrical sizes $ka\in\{50,500\}$, see App.~\ref{S:RadiationModes}. The spherical shell has surface resistivity $R_\mrm{s}$, and the spherical ball has resistivity $\rhor$. Both the surface and volumetric objects exhibit similar normalized radiation modes, each showing a cutoff around $n\approx \Na=2(ka)^2$. The cutoff becomes tighter with increasing electrical size, with the $ka=500$ case approaching a step function for the spherical ball.

Alternatively, instead of using a threshold for efficiency, the effective NDoF~\cite{Shiu+etal2000,Yuan+etal2022} can be utilized, which, when expressed in terms of radiation mode efficiency, is given by
\begin{equation}
	N_{\mrm{e}} 
	= \frac{(\Tr\M{H}\M{H}^{\herm})^2}{\norm{\M{H}\M{H}^{\herm}}_\T{F}^2}
	= \frac{\left(\sum_{n=1}^{\infty}\nu_n\right)^2}{\sum_{n=1}^{\infty}\nu_n^2}.
\label{eq:eNDoF}
\end{equation}
Fig.~\ref{fig:NDoFsphN_tols_ka} compares the NDoF $\Nr$ based on $50\%$ efficiency, as
in Fig.~\ref{fig:NDoFsph_tols_ka}, with the asymptotic effective NDoF $N_\T{e}$ in~\eqref{eq:eNDoF}. The depicted NDoFs are normalized with the estimate $\Na=2(ka)^2$ for the spherical region. The NDoF based on $\radm_n\geq 1$ and the NDoF ~\eqref{eq:eNDoF} agree well but the $\radm_n\geq 1$ formulation oscillates due to the onset of modes at specific frequencies. The NDoFs for electrically smaller regions are above the $\Na=2(ka)^2$ estimate~\eqref{eq:NDoFsph} but approaches it for larger sizes. 

\begin{figure}[t]%
\includegraphics[width=\columnwidth]{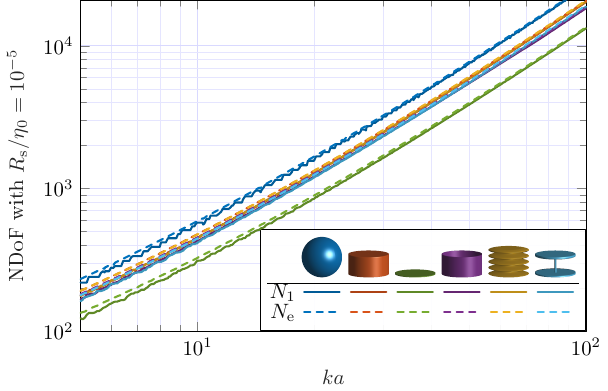}%
\caption{NDoF $\Nr$ from radiation modes $\radm_n\geq 1$ in~\eqref{eq:radmeig} and $N_\T{e}$ in~\eqref{eq:eNDoF} for the six objects in Tab.~\ref{tab:Area} with surface resistivity $R_\mrm{s}=10^{-5}\eta_0$.}%
\label{fig:NDoFradmod_ka}%
\end{figure}

Radiation modes and hence NDoF defined by~\eqref{eq:radmeig} or~\eqref{eq:eNDoF} can be calculated for arbitrary shapes and material parameters. Fig.~\ref{fig:NDoFradmod_ka} shows an example of NDoF for the six different shapes in Tab.~\ref{tab:Area} modelled by a surface resistivity $R_{\mrm{s}}=10^{-5}\eta_0$. It is observed that the NDoF increase with the electrical size $ka$, with the spherical shell exhibiting the highest NDoF, followed by the solid cylinder and the corrugated cylinder, which have slightly higher NDoF than the open cylinder and connected discs. The single disc exhibits the lowest NDoF. The effective NDoF~\eqref{eq:eNDoF} produces similar results but with fewer oscillations. 
This paper demonstrates that the asymptotic number of NDoF is proportional to the electrical size of the region's average shadow area $\ave{\As}$ in Tab.~\ref{tab:Area}, see also~\cite{Gustafsson+Lundgren2024} for the corresponding number of significant characteristic modes.

\begin{table}[t]
\caption{Average shadow area $\ave{A_\mrm{s}}$, surface area $A$, and height to radius ratio $h/r$ for a sphere, cylinder, disc, open cylinder, corrugated cylinder, and connected discs. Objects are circumscribed by a sphere with radius $a$.}
{\centering	
\begin{tabular}{|c|cccccc|} \hline
& \cincludegraphics[width=0.7cm]{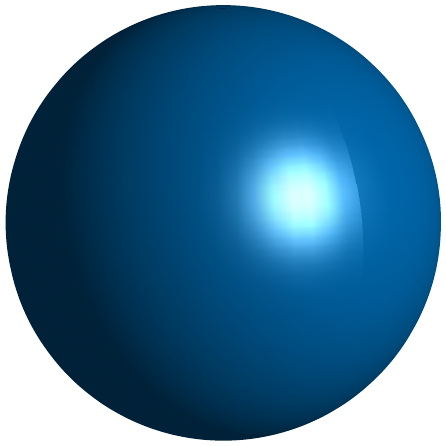} 
& \cincludegraphics[width=0.7cm]{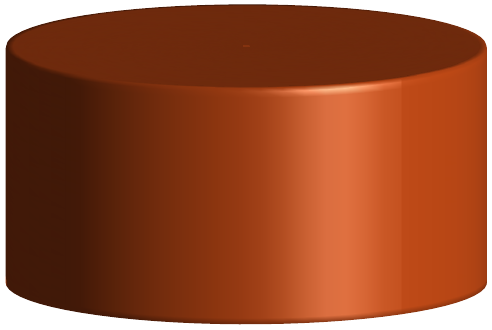} 
& \cincludegraphics[width=0.7cm]{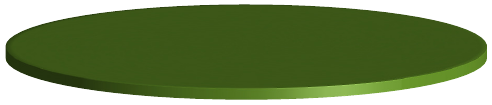} 
& \cincludegraphics[width=0.7cm]{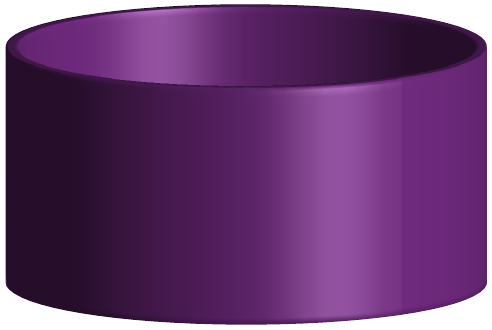} 
& \cincludegraphics[width=0.7cm]{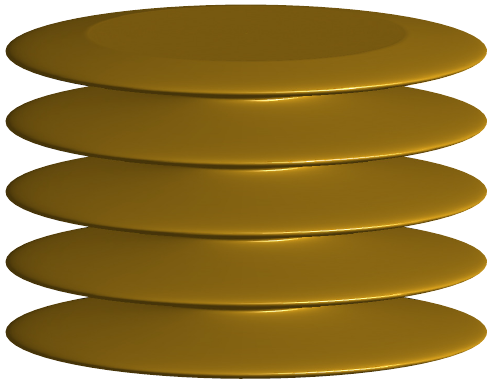} 
& \cincludegraphics[width=0.7cm]{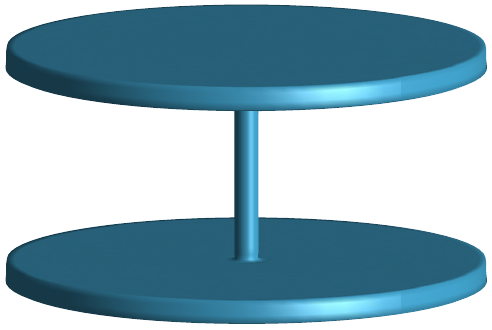} 
\\ \hline
$\ave{A_{\mrm{s}}}/a^2$ & 3.14 & 2.53 & 1.65 & 2.21 & 2.45 & 2.19\\
$ A/a^2$ & 12.6 & 10.1 & 6.59 & 10.3 & 21.9 & 11.3\\ 
$ A/\ave{A_{\mrm{s}}}$ & 4.00 & 4.00 & 4.00 & 4.65 & 8.92 & 5.15\\ 
$ h/r$ & 2.00 & 1.00 & 0.05 & 1.00 & 1.33 & 1.00\\ 
\hline
\end{tabular}
\par}
\label{tab:Area}
\end{table}

\section{NDoF and shadow area}\label{S:NDoFshadowarea}
Determination of NDoF from the number of radiation modes~\eqref{eq:radmeig} greater than unity $\Nr$ or from the effective NDoF~\eqref{eq:eNDoF} $N_\T{e}$ is easily achieved numerically for arbitrary shaped objects and material parameters~\cite{Ehrenborg+Gustafsson2020}, see Fig.~\ref{fig:NDoFradmod_ka}. This numerical evaluation is complemented by an analytical examination, providing valuable physical insight and understanding. Specifically, it demonstrates that the radiating NDoF in the electrically large limit is proportional to the shadow area measured in squared wavelengths~\cite{Gustafsson+Lundgren2024}.

This proportionality can be derived using scattering or antenna theory. Here, employing antenna theory, the maximal effective area is first expressed in terms of radiation modes and subsequently related to the cross-sectional area of the object. The maximal partial effective area $A_{\mrm{eff}}=\lambda^2 G/(4\pi)$, and similarly the partial gain $G$ in a direction $\UV{k}$ and polarization $\UV{e}$ are determined from the solution of~\cite{Gustafsson+Capek2019}
\begin{equation}
\begin{aligned}
& \maximize && A_{\mrm{eff}}=\lambda^2\M{I}^{\herm}\M{F}^{\herm}\M{F}\M{I}  \\
& \subto && \M{I}^{\herm}\M{R}\M{I}=1,
\end{aligned}
\label{eq:MaxAeff}
\end{equation}
where $\M{F}\M{I}$ is proportional to the $\UV{e}$-component of the far field in the direction $\UV{k}$, and $\M{I}^{\herm}\M{F}^{\herm}\M{F}\M{I}$ to the corresponding partial radiation intensity~\cite{Gustafsson+Capek2019}. The solution of~\eqref{eq:MaxAeff} is
\begin{equation}
    \max\Aeff(\UV{k},\UV{e}) 
    = \lambda^2\M{F}^{\herm}\M{R}^{-1}\M{F}
\label{eq:MaxEff2}
\end{equation}
and taking the average of $\max\Aeff(\UV{k},\UV{e})$ over all directions $\UV{k}$ and polarizations $\UV{e}$
\begin{equation}
    \ave{\max\Aeff} = \frac{1}{8\pi^2}\int_{2\pi}\int_{4\pi}
    \max\Aeff(\UV{k},\UV{e})
    \diff\Omega_{\UV{k}}\diff\Omega_{\UV{e}}
    \label{eq:average}
\end{equation}
and similarly for the average maximal partial gain $\ave{\max G}=4\pi\ave{\max\Aeff}/\lambda^2$.
By diagonalizing the optimization problem~\eqref{eq:MaxAeff} using the radiation modes~\eqref{eq:radmeig} and performing the directional and polarization averaging~\eqref{eq:average} detailed in App.~\ref{S:Averages}, we derive a straightforward expression for the average maximal effective area, formulated in terms of the radiation modes
\begin{equation}
	\ave{\max\Aeff} 
	=\frac{\lambda^2}{8\pi}\sum_{n=1}^{\infty}\nu_n
	=\frac{\lambda^2}{8\pi}\sum_{n=1}^{\infty}\frac{\radm_n}{1+\radm_n}.
\label{eq:maxavAff}
\end{equation}

To estimate the number of modes, we utilize that radiation modes decay rapidly after a finite number of modes, similar to the estimate for spherical modes $L\approx ka$ in~\eqref{eq:NDoFsph} shown in Fig.~\ref{fig:radmodessph}. This assumption is further supported by a sum rule identity for radiation modes of homogeneous objects~\cite{Gustafsson+etal2020}
\begin{equation}
	\sum_{n=1}^{\infty}\radm_n 
	=\frac{2\pi\eta_0 V}{\lambda^2\rhor}
\label{eq:RadmSumRule}
\end{equation}
which demonstrates that radiation modes $\radm_n$ decay, as the sum of all radiation modes is fixed and proportional to the volume and inversely proportional to the resistivity. 

For typical surface resistivities of metals, such as $R_\mrm{s}\approx 0.01\unit{\Omega/\square}$ for copper around $1\unit{GHz}$, the radiation modes approximately divide into two groups: those with $\radm_n\gg 1$ and those with $\radm_n\ll 1$, see Fig.~\ref{fig:radmodessph}, resulting in efficiencies $\nu_n$ according to
\begin{equation}
 \nu_n=\frac{\radm_n}{1+\radm_n} \approx 
	\begin{cases}
			1 & n < \Nr \\
			0 & n > \Nr, 
	\end{cases}
\label{eq:radmeff}
\end{equation}
where $\Nr$ denotes the NDoF for the given shape and frequency. The distribution is continuous, which means that a few modes have efficiencies between 1 and 0. The transition from large to small modal efficiencies decreases with increasing electrical size, see Fig.~\ref{fig:radmodessph}. 

Inserting~\eqref{eq:radmeff} into~\eqref{eq:maxavAff} produces the estimate
\begin{equation}
	\ave{\max\Aeff} 
	\approx
	\frac{\lambda^2}{8\pi}
	\Nr.
\label{eq:MaxAeffNDoF}
\end{equation}
Rewriting the partial effective area in the partial gain, $G=4\pi A_{\mrm{eff}}/\lambda^2$, suggest that the NDoF can alternatively be expressed as
\begin{equation}
	\Nr\approx 2\ave{\max G} 
	=\frac{1}{2}\sum\frac{\radm_n}{1+\radm_n},
\label{eq:NDoFMaxGain}
\end{equation}
where $\ave{\max G}$ denotes the average maximum partial gain ~\cite{Gustafsson+Capek2019}. This expression resembles the NDoF in~\cite{Kildal+etal2017} based on the maximal directivity in a fixed direction for a spherical region, but is here generalized to arbitrary shaped lossy objects. 

The radiation modes are related to the geometrical structure by using that the maximal effective area in a direction $\UV{k}$ approaches the shadow area (geometrical cross section for convex shapes), $\As(\UV{k})$, in the electrically large limit~\cite{Gustafsson+Capek2019} $	\max\Aeff(\UV{k}) \to \As(\UV{k})$ and similarly for the average over polarizations and directions
\begin{equation}
\ave{\max\Aeff} \to \ave{\As}	
=\frac{1}{4\pi}\int_{4\pi}\As(\UV{k})\diff\Omega_{\UV{k}}    
\label{eq:Aeff2As}
\end{equation}
as $ka\to\infty$.
This connects the radiation modes and geometrical properties of the region $\reg$ 
with the maximal effective area~\eqref{eq:Aeff2As} expressed solely in geometrical parameters producing the asymptotic NDoF
\begin{equation}
	\Nr \approx \Na
    =\frac{8\pi \ave{A_\T{s}}}{\lambda^2}	
	=\frac{2|A_\T{s}|}{\lambda^2}
	\quad\text{as } ka\to\infty,
\label{eq:NDoF_As}
\end{equation}
where $|A_\T{s}|=4\pi\ave{A_\T{s}}$ denotes the total shadow area.
This demonstrates that the NDoF is approximately twice to the total shadow area measured in squared wavelengths, where 2 stems from the two polarizations. Additionally, the NDoF is twice the number of significant characteristic modes~\cite{Gustafsson+Lundgren2024}.

\begin{figure}
    \centering
    \begin{tikzpicture}
        \node at (0,0) {\includegraphics[width=0.9\columnwidth]{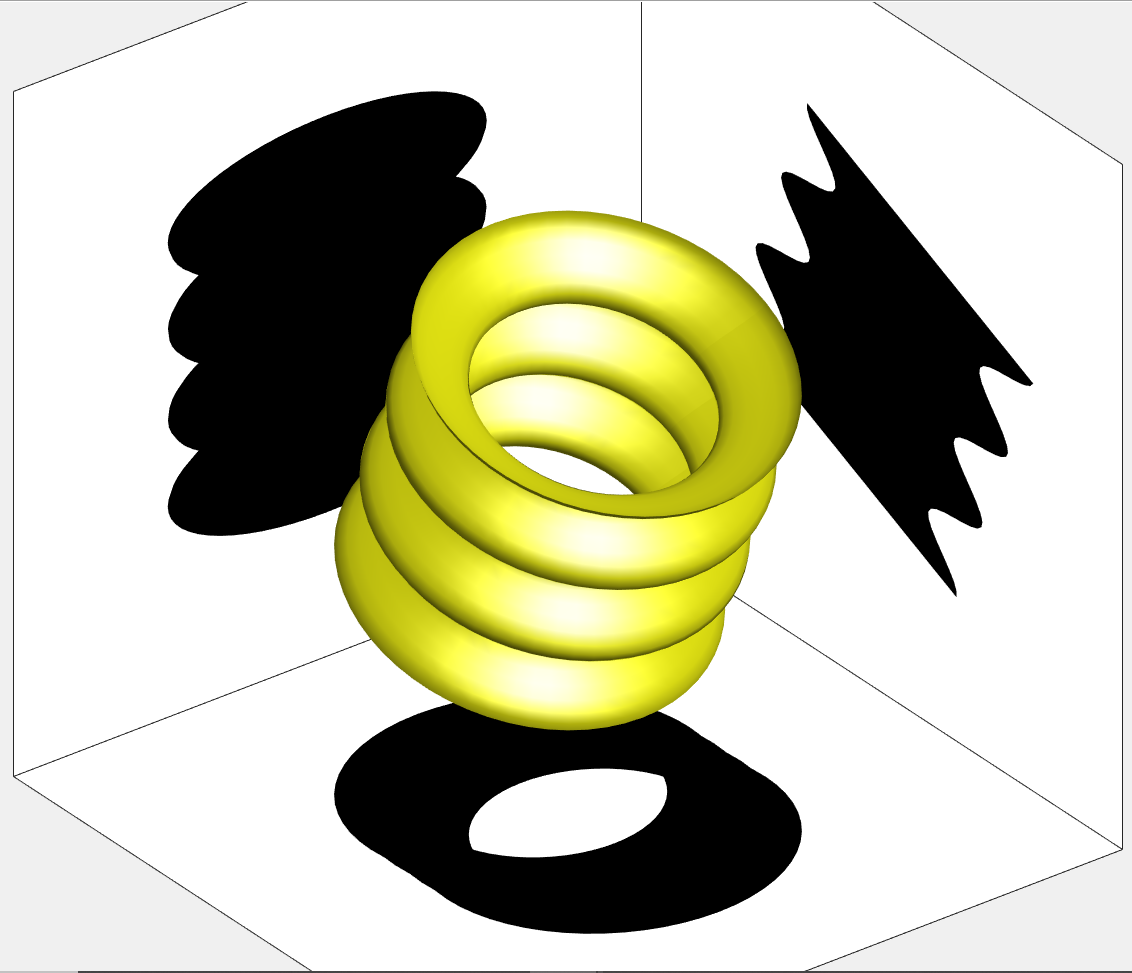}};
        \node[white] at (2.2,1) {$A_\T{s}(\UV{x})$};
        \node[white] at (-2,1.8) {$A_\T{s}(\UV{y})$};
        \node[white] at (-1,-2) {$A_\T{s}(\UV{z})$};
    \end{tikzpicture}
    \caption{Illustration of the shadow area $A_\T{s}(\UV{k})$ of a tilted corrugated cylinder for illumination directions $\UV{x}$, $\UV{y}$, and $\UV{z}$, as projected onto a plane perpendicular to each respective illumination direction.}
    \label{fig:ShadowAreaXYZ}
\end{figure}

The shadow area, $A_\T{s}(\UV{k})$, is determined by illuminating the (opaque) region in the direction $\UV{k}$ and projecting the resulting shadow onto a plane perpendicular to this direction, as shown in Fig.~\ref{fig:ShadowAreaXYZ}. For convex shapes, the shadow area is equivalent to the maximum cross-sectional area in the given direction. Note that the shadow area is symmetric $A_\T{s}(\UV{k})=A_\T{s}(-\UV{k})$.

The NDoF and shadow area are first analyzed for a spherical region with radius $a$. Due to the spherical symmetry, the shadow area is independent of the illumination direction, $A_\T{s}(\UV{k}) = \pi a^2$, corresponding to the cross-sectional area. Consequently, the average shadow area~\eqref{eq:Aeff2As} over all directions is also $\ave{A_\T{s}} = \pi a^2$. The NDoF, given by~\eqref{eq:NDoF_As}, is $2(ka)^2$, consistent with the classical result in~\eqref{eq:NDoFsph}.  
The corresponding average maximal gain, derived from~\eqref{eq:NDoFMaxGain}, is $(ka)^2$, which matches the leading term in Harrington's gain estimate~\cite{Harrington1960}.

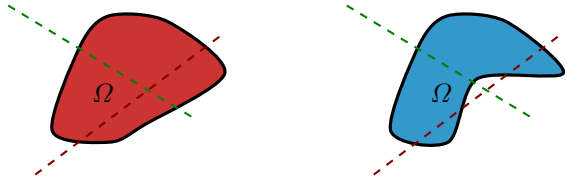
\begin{figure}%
{\centering
\begin{tikzpicture}[scale=0.9,>= stealth]
	\begin{scope}[xshift=0cm]
	\draw [fill=preg,very thick] plot [smooth cycle,scale=0.85] coordinates {(-1,-1) (0,-1.2) (0.6,-0.9) (2,0) (1,0.9) (0,1) (-0.5,0.5)};    
	\node at (-0.1,-0.3) {$\reg$};	
\draw[dashed,thick,dgreen] (-1.5,1) --  (1.3,-0.7);	
\draw[dashed,thick,dred] (-1.1,-1.5) --  (1.7,0.6);	
	\end{scope}	
	\begin{scope}[xshift=5cm]
	\draw [fill=breg,very thick] plot [smooth cycle,scale=0.85] coordinates {(-1,-1) (0,-1.2) (0.5,-0.1) (2,0) (1,0.9) (0,1) (-0.5,0.5)};    
	\node at (-0.1,-0.3) {$\reg$};	
\draw[dashed,thick,dgreen] (-1.5,1) --  (1.3,-0.7);	
\draw[dashed,thick,dred] (-1.1,-1.5) --  (1.7,0.6);	
	\end{scope}	
\end{tikzpicture}
\par}
\caption{Shadow area for a region $\reg\in\R^3$. (left) convex object having  rays intersecting the boundary $\breg$ twice. (right) non-convex object with some rays intersecting the boundary more than twice.}%
\label{fig:ShadowArea}%
\end{figure}

For a spherical region, the total surface area, $A = 4\pi a^2$, is exactly four times the average shadow area, $A=4\ave{A_\T{s}}$. This relationship is a general result for convex shapes, originally derived by Cauchy~\cite{Vouk1948}
\begin{equation}
	\ave{\As}=A/4.
\label{eq:ShadowAreaConvex}
\end{equation}
This identity follows from the principle that each ray intersecting the object also intersects its boundary twice~\cite{Vouk1948}, as illustrated in Fig.~\ref{fig:ShadowArea}.
For non-convex objects, rays might intersect the boundary more than twice. This can be interpreted as shadowing of some regions, or equivalently, that some boundary points cannot radiate undestructively over $2\pi$ steradian to the farfield. For convex objects with surface area $A$, the NDoF~\eqref{eq:NDoF_As} reduces to     
\begin{equation}
\Na\approx  
\frac{8\pi\ave{A_{\mrm{s}}}}{\lambda^2}
\stackrel{\substack{\mrm{convex}\\\mrm{object}}}{=}
\frac{2\pi A}{\lambda^2}
=\frac{k^2 A}{2\pi}
\quad\text{as }
ka\to\infty.
\label{eq:NDoFconvex}
\end{equation}
This estimate agree with the $L=ka$ cutoff for spherical regions~\eqref{eq:NDoFsph} and Weyl's law for convex shaped regions~\eqref{eq:Weyl2D} but differ for non-convex shapes, see Tab.~\ref{tab:Area}.

\begin{figure}[t]%
\includegraphics[width=\columnwidth]{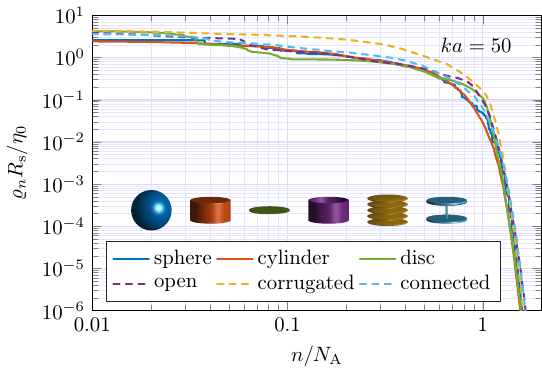}%
\caption{Normalized radiation modes $\radm_n$ for the six objects in Tab.~\ref{tab:Area} of the electrical size $ka=50$ plotted versus the normalized mode index $n/\Na$~\eqref{eq:NDoF_As}.}%
\label{fig:radmodes}%
\end{figure}

Radiation modes for six different objects are depicted in Fig.~\ref{fig:radmodes}. The radiation mode eigenvalues $\radm_n$ are normalized with the surface resistivity $R_{\mrm{s}}$, and the mode index $n$ is scaled with the NDoF, $\Na$, based on the shadow area~\eqref{eq:NDoF_As}, as shown in Tab.~\ref{tab:Area}. The $\Na$ dominant normalized radiation modes have approximately unit magnitude, after which they rapidly decay.

\begin{figure}[t]%
\includegraphics[width=\columnwidth]{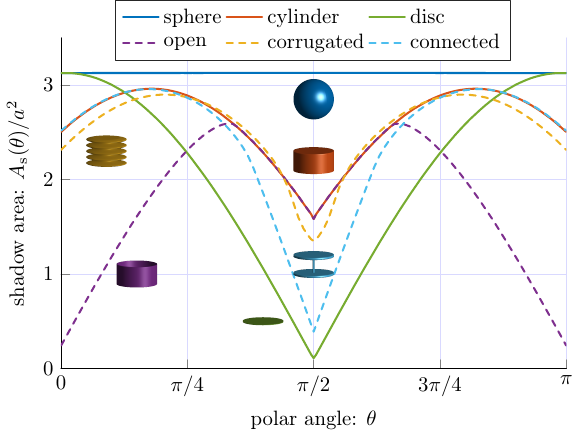}%
\caption{Shadow areas for the six objects in Tab.~\ref{tab:Area} as function of the polar angle $\theta$. The shadow areas are normalized by the smallest object's radius $a^2$.}%
\label{fig:shadowarea}%
\end{figure}

Shadow areas for the objects in Tab.~\ref{tab:Area} are independent of the azimuthal coordinate $\phi$ and plotted as function of the polar angle $\theta$ in Fig.~\ref{fig:shadowarea}. The sphere has $A_{\mrm{s}}(\theta)=\pi a^2$ for all directions. The disc is also approximately $A_{\mrm{s}}(0)\approx\pi a^2$ in the normal direction but much smaller seen from the side $A_{\mrm{s}}(\theta)\approx \pi a^2\cos\theta$. The solid cylinder and connected discs coincide for illuminations from above, $\theta\in[0,0.2 \pi]$, but start to differ for illuminations from the side. The opposite holds for the cylinder compared with the open cylinder, where the opening is observed from the above but not from the side. The corrugated cylinder has similar cross-sectional area as the cylinder but differ for illuminations from the side where the corrugations are visible, \cf~Fig.~\ref{fig:ShadowAreaXYZ}. The difference at $\theta=0$ is due to the different aspect ratio, see Tab.~\ref{tab:Area}. The shadow areas are bounded by $\pi a^2$ which is reached for a sphere for all directions but also approximately for thin discs. The minimal values can approach zero, as observed for the open cylinder at $\theta=0$ and disc at $\theta=\pi/2$. The NDoF depends on the total shadow area~\eqref{eq:NDoF_As}, which is determined by integration of the curves in Fig.~\ref{fig:shadowarea} weighted by $\sin\theta$.

The NDoF~\eqref{eq:NDoF_As} depends on the material properties such as surface resistivity $R_{\mrm{s}}$ or bulk resistivity $\rhor$. Metals typically have values $R_{\mrm{s}}/\eta_0$ and $k\rhor/\eta_0$ below $10^{-4}$ producing a distinct difference between efficient and inefficient modes as exemplified in Fig.~\ref{fig:radmodes} and Fig.~\ref{fig:radmodessph}. Highly resistive materials have much larger resistivity, giving fewer efficient modes and hence less DoF. Materials have here for simplicity been treated as homogeneous, but the formulation applies equally well for inhomogeneous regions. 

The constraint based on material losses such as in~\eqref{eq:Rayleighradmod} can alternatively be interpreted as restrictions of the amplitude of the current density. This is \eg seen from the loss matrix of a homogeneous object, where the loss matrix separates into the resistance times the Gram matrix $\M{R}_{\rho}=\rhor\M{\Psi}$. Here, the least-squares norm of the current density is approximately
\begin{equation}
	\int_\reg|\V{J}(\V{r})|^2\diffV \approx \M{I}^{\herm}\M{\Psi}\M{I}.
\label{eq:JL2}
\end{equation}   
The normalized parameters $R_{\mrm{s}}/\eta_0$ and $k\rhor/\eta_0$ can hence be interpreted as a weight for the current norm.

\section{Electric and magnetic currents}\label{S:EMcurrents}
The NDoFs determined in Secs.~\ref{S:capacity} and~\ref{S:NDoFshadowarea} are based on the assumption of lossy non-magnetic materials or similarly, a restriction on the amplitude of the electric current $\V{J}$. However, magnetic materials and magnetic currents $\V{M}$ are also important to consider in fully understanding the NDoF of a region. Linear magnetic materials are typically lossy and can be included in the analysis using volumetric magnetic contrast currents together with a magnetic loss matrix $\M{R}_{\rho,\T{m}}$. Magnetic surface currents are mainly important for their use as equivalent currents used to describe the field outside a region~\cite{Kristensson2016}.

Allowing for magnetic currents increases the NDoF compared with the solely electric case. This is particularly evident for electrical sizes where the geometrical structure is not resolved by the wavelength. 
NDoF of infinitely thin sheets such as planar regions do not directly follow from Weyl's law, as they do not have an inner region. Considering \eg a planar region with only electric currents enforces a symmetry of the radiated field in the up and down directions reducing the NDoF~\cite{Ehrenborg+Gustafsson2020}. 
This symmetry is broken by allowing both electric and magnetic currents on the thin sheet, and hence making the structure behave as a volumetric region. 

\begin{figure}[t]%
{\centering
\includegraphics[width=\columnwidth]{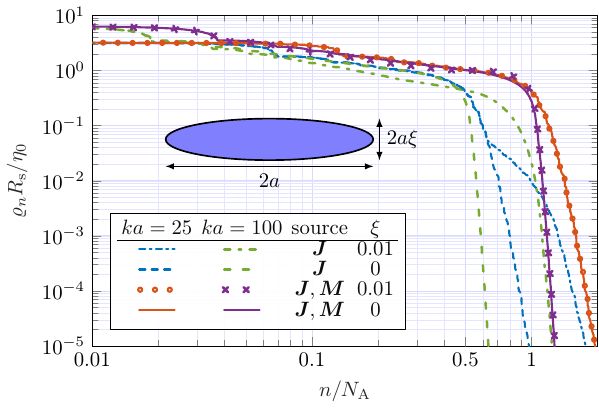}%
\par}
\caption{Normalized radiation modes for an oblate spheroid ($\xi=0.01$) and disc ($\xi=0$) using only electric current density $\V{J}$ or both electric and magnetic $\V{M}$ current densities.}%
\label{fig:oblatevsdisc}%
\end{figure}

Assuming magnetic losses similar to the electric loss matrix $\M{R}_\rho$ in~\eqref{eq:CapRadm} does not affect the asymptotic NDoF~\eqref{eq:NDoF_As} given by the total shadow area. Numerical evaluation of the NDoF depends however on the loss parameters and including magnetic currents increases NDoF. Fig.~\ref{fig:oblatevsdisc} shows the radiation modes for oblate spheroidals with aspect rations $\xi=0.01$ and $\xi=0$. The magnetic surface losses are set to correspond to the electric losses. 

Considering first the case with both electric, $\V{J}$, and magnetic currents, $\V{M}$, shown by solid curves and markers in Fig.~\ref{fig:oblatevsdisc}. The radiation modes for oblate spheroids with $\xi=0.01$ and $\xi=0$ (infinitely thin disc) overlap. This is due to the negligible thickness of the $\xi=0.01$ oblate spheroid. The radiation modes $\radm_n$ have a cutoff around the NDoF estimate~\eqref{eq:NDoF_As} based on the average shadow area $\ave{\As}=\pi a^2/2+\pi a^2\xi^2/(4 e)\ln((1+e)/(1-e))$ with $e^2=1-\xi^2$, which approaches $\pi a^2/2$ as $\xi\to 0$. The slope of $\radm_n$ for $n>\Na$ becomes steeper as $ka$ increases, similar to the results for the sphere in Fig.~\ref{fig:radmodessph}. Here, we also note that the disc $\xi=0$ is treated as having a top and bottom, which is equivalent to using half the resistivity for the disc.

The purely electric case ($\V{M}=\V{0}$) is more complex and shows a clear difference between the disc ($\xi=0$) and oblate spheroid ($\xi=0.01$). Radiation modes for the disc show a cutoff around $n/\Na=0.5$, indicating a halving of the NDoF. This originates from the symmetric radiation of the electric currents in the disc. The finite thickness of the oblate spheroid breaks this symmetry and increases the NDoF. The radiation modes are similar up to $n/\Na\approx 0.5$ but differ for larger mode indices $n$. Radiation modes for the disc decay rapidly after the cutoff around $n/\Na\approx 0.5$ with a slope increasing with $ka$. Radiation modes for the oblate spheroid of the smaller size $ka=25$ also have a cutoff around $n/\Na\approx 0.5$ but then deviate from the disc modes and start to approach the traces for the $\V{J}$ and $\V{M}$ cases. The $n/\Na\approx 0.5$ cutoff vanishes for the electrically larger case $ka=100$, which instead shows a $n/\Na\approx 1$ cutoff as for the $\V{J}$ and $\V{M}$ cases. Note that the electric thickness of the oblate spheroid is $2ka\xi\in\{0.5, 2\}$ for the $ka\in\{25,100\}$ cases.
      
Needle shaped prolate spheroidals are similarly one dimensional, showing approximately a different scaling as evident from Weyl's law~\eqref{eq:Weyl} before their thickness is resolved. 
			
\begin{figure}%
{\centering
\includegraphics[width=\columnwidth]{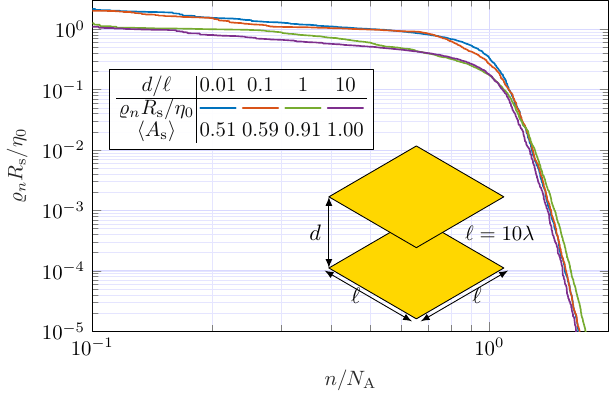}%
\par}
\caption{Radiation modes for two square plates with side lengths $\ell$ separated a distance $d$ at electrical sizes $\ell=10\lambda$. Electric and magnetic currents are used to emulate volumetric regions. The mode index is normalized with the asymptotic estimate~\eqref{eq:NDoF_As} based on the average shadow area $\ave{\As}$.}%
\label{fig:RadmTwoRec}%
\end{figure}

\begin{figure}%
{\centering
\includegraphics[width=\columnwidth]{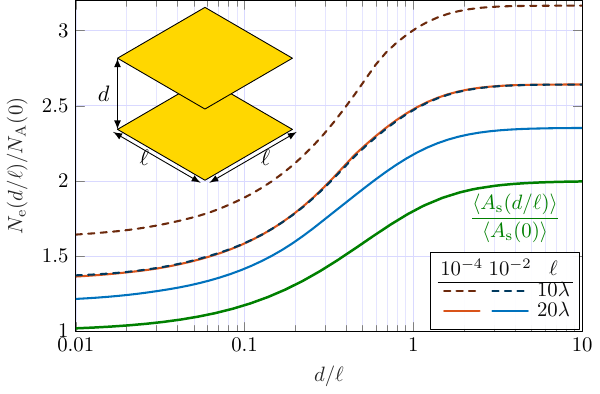}%
\par}
\caption{Effective degrees of freedom~\eqref{eq:eNDoF} for two square plates with side lengths $\ell$ separated a distance $d$ at electrical sizes $\ell/\lambda\in[10,20]$. Electric and magnetic currents are used to emulate volumetric regions. The eNDoF is normalized with the asymptotic estimate~\eqref{eq:NDoF_As} evaluated for a single plate. The results are compared with the average shadow area for distances $d/\ell$ normalized with the average shadow area for a single plate.}%
\label{fig:TwoRec}%
\end{figure}

The presented derivation of the NDoFs remain valid for non-connected and multiple objects, as demonstrated by the two square plates in Figs.~\ref{fig:RadmTwoRec} and~\ref{fig:TwoRec}. Fig.~\ref{fig:RadmTwoRec} illustrates the radiation modes $\radm_n$ for two plates separated by a distance $d/\ell\in\{0.01,0.1,1,10\}$. The modal index $n$ is normalized with the asymptotic NDoF $\Na$ based on the average shadow area $\ave{\As}$. Notably, the radiation modes $\radm_n$ exhibit a cutoff around $\Na$. The shadow area can be numerically evaluated by considering the projected area of the two squares in a plane parallel to the plates. This resulting shadow area, plotted versus the distance $d$, is illustrated in Fig.~\ref{fig:TwoRec} by the curve $\ave{\As(d/\ell)}/\ave{\As(0)}$, where $\ave{\As(0)}=\ell^2/2$ represents the average shadow area for a single square, as shown in equation~\eqref{eq:ShadowAreaConvex}. As the distance increases, the average shadow area also increases and approaches $2\ave{\As(0)}$ as $d/\ell\to\infty$, corresponding to the NDoF of a plate with an area of $2\ell^2$.

The effective NDoFs~\eqref{eq:eNDoF} for $\ell/\lambda\in\{10,20\}$ are depicted in Fig.~\ref{fig:TwoRec}. The curves align with the trend of the asymptotic value~\eqref{eq:NDoF_As} derived from the average shadow area $\ave{\As}(d/\lambda)$. Moreover, they converge towards the asymptotic result as the electrical size increases, as observed for the $\ell/\lambda\in\{10,20\}$ cases.
 
\section{Inverse source problem}\label{S:InverseProblems}

\begin{figure}%
{\centering
\begin{tikzpicture}[scale=0.9,>= stealth]
	\begin{scope}[xshift=5cm]
	\draw [fill=breg,very thick] plot [smooth cycle,scale=0.85] coordinates {(-1,-1) (0,-1.2) (0.5,-0.1) (2,0) (1,0.9) (0,1) (-0.5,0.5)};    
	\node at (0.5,-0.3) {$\reg$};	
	\node at (0.5,1.2) {$\vec{J}(\rv)$};	
\draw[->,thick,decorate,decoration=snake] (-1,0) -- node[sloped,below] {$\M{f}$} (-1.7,1);	
\draw[dashed,dred] (0,0) circle(2cm);
	\node at (-1.5,1.5) {Rx};	
	\node at (-2,2) {b)};	
	\end{scope}	
\end{tikzpicture}
\vspace{-1mm}
\par}
\caption{Inverse source problem for a region with surface $\reg$ with measurements in the far field or on a circumscribing sphere.}%
\label{fig:Invsource}%
\end{figure}

The concept of NDoF for radiating systems, as analyzed in Sec.~\ref{S:NDoFshadowarea}, also applies to inverse source problems. Consider a region composed of some linear material. By employing the equivalence principle, the radiated field can be represented by electric $\V{J}$ and magnetic $\V{M}$ surface current densities on a surface $\reg$ surrounding the region~\cite{vanBladel2007}. For simplicity, we let this surface coincide with the boundary $\reg$ of the region. With all sources inside $\reg$, Love's theorem or the null field condition can be used to relate the electric and magnetic currents in the inverse source problem~\cite{Persson+Gustafsson2005,AraqueQuijano+Vecchi2010}. 

Considering for simplicity electrical currents on a PEC object. Expanding the measured field in spherical waves with expansion coefficients $f_n$  and adding noise to the measured field (expansion coefficients). This results in the measurement model
\begin{equation}
	\M{f} = -\M{U}\M{I}+\M{n},
\label{eq:InvSourceModel}
\end{equation}
which resembles the communication model in Sec~\ref{S:capacity}. In inverse source problems, the source current $\M{I}$ is estimated from the observed field $\M{f}$. The system~\eqref{eq:InvSourceModel} is not invertible, and it is common to reformulate the system as an optimization problem or to use an SVD for a regularized solution~\cite{Hansen2010}.

Regularizing the inverse source problem~\eqref{eq:InvSourceModel} by penalizing the norm of the current density~\cite{Hansen2010}, \eg
\begin{multline}
	\min_{\M{I}}\norm{\M{U}\M{I}+\M{f}}^2 + \delta\eta_0\norm{\M{I}}_{\reg}^2 \\
	=\min_{\M{I}} \M{I}^{\herm}\M{R}_0\M{I}+2\Re\{\M{f}^{\herm}\M{U}\M{I}\}
	+|\M{f}|^2 + \delta\eta_0\M{I}^{\herm}\M{\Psi}\M{I},	
\label{eq:InvSourceOpt}
\end{multline}
where the induced metric~\eqref{eq:JL2} from $\reg$ is used for the current. Here, the regularization term $\delta\eta_0\M{I}^{\herm}\M{\Psi}\M{I}$ has the same form as for the dissipated power in the material from the material matrix $\M{R}_{\rho}$ in~\eqref{eq:RadmEff}, \ie the regularization parameter $\delta$ mimics the normalized surface resistivity $R_{\mrm{s}}/\eta_0$. This similarity reveals a connection between the physical bounds formulated in currents~\cite{Gustafsson+Nordebo2013} and inverse source problems.

Expanding the current
\begin{equation}
	\M{I} = \sum_n I_n\M{I}_n
\label{eq:CurrentiInRadmodes}
\end{equation}
in radiation modes~\eqref{eq:radmeig} defined by 
using $\M{R}_{\rho}=\eta_0\M{\Psi}$ to diagonalize~\eqref{eq:InvSourceOpt} 
\begin{equation}
	\min_{I_n} \sum_n (\radm_n+\delta)|I_n|^2+2\Re\{f_n^{\ast}I_n\sqrt{\radm_n}\} + |f_n|^2
\label{eq:InvSourceRadm}
\end{equation}
with the solution 
\begin{equation}
	I_n
	=-\frac{\sqrt{\radm_n}}{\delta+\radm_n}{f}_n.
\label{eq:invsourceOpt}
\end{equation}
The corresponding solution using an SVD is
\begin{equation}
	I_n = -f_n/\sqrt{\radm_n}.
\label{eq:invsourceSVDradm}
\end{equation}
The solutions~\eqref{eq:invsourceSVDradm} and~\eqref{eq:invsourceOpt} are similar for efficient radiation modes $\radm_n\gg 1$ but differ for inefficient modes $\radm_n\ll 1$. Inefficient radiation modes make the noise in~\eqref{eq:InvSourceModel} blow up for~\eqref{eq:invsourceSVDradm} unless they are removed in the pseudo inverse related to the SVD. The (Tikhonov) regularized solution has a smoother transition where the impact of inefficient modes decreases as $\sqrt{\radm_n}$. The efficient modes are also the ones contributing to the NDoF.   

These simple problems demonstrate the utility of NDoF derived from radiation modes and, asymptotically, from the shadow area in understanding inverse source problems. Employing the asymptotic NDoF offers a straightforward estimation of resolution in such problems. For instance, if the surface area is $|\reg|$, then there are approximately $2\pi |\reg|/\lambda^2$ DoFs for the current density on $\reg$ according to~\eqref{eq:Weyl2D}, while the measured data contributes approximately $8\pi \ave{\As}/\lambda^2$ DoFs based on the average shadow area $\ave{\As}$.

\begin{figure}%
{\centering
\includegraphics[width=0.4\columnwidth]{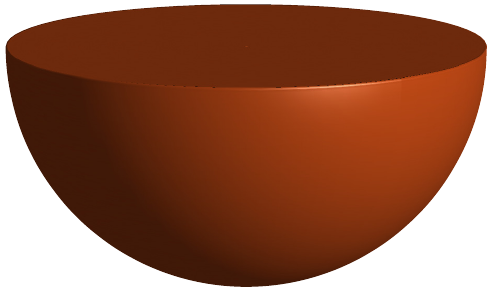}%
\includegraphics[width=0.4\columnwidth]{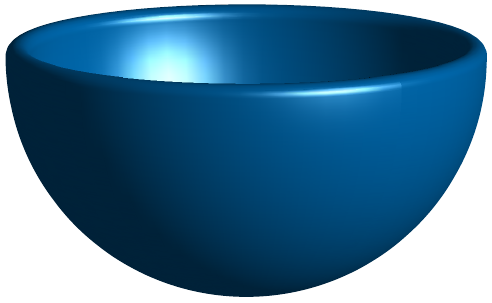}%
\par}
\caption{Inverse source problems for a hemisphere (convex) and a bowl (non-convex). Both object have radius $a$.}%
\label{fig:halfsphere}%
\end{figure}

As an example, consider a hemisphere (half ball) with radius $a$ and a half spherical shell with radius $a$ and thickness $a/10$, as depicted in Fig.~\ref{fig:halfsphere}. The hemisphere is convex and has a surface area $A=3\pi a^2$ and an average shadow area $\ave{\As}=A/4=3\pi a^2/4$. The half spherical shell has an identical shadow area and an approximate surface area $A\approx 4\pi a^2$ (assuming thin walls). This implies that the radiating NDoF asymptotically are the same for the two objects, but the non-convex bowl has a larger surface area over which the current is reconstructed. This suggests that the same NDoF are used to reconstruct the current density over different surface areas.
As a consequence, this reduces the (average) resolution from approximately $\lambda/1.8$ for the hemisphere with $75\%$ to approximately $\lambda/1.5$ for the bowl.

\section{Conclusions}\label{S:conclusion}
We present a method for determining the NDoF for a radiating system based on radiation modes. Asymptotically, the NDoF approaches the total shadow area measured in squared wavelengths. For convex regions, the NDoF matches the results derived from Weyl's law. The findings extend the known expressions to arbitrary shapes, encompassing non-convex and non-connected regions. Numerical results complementing the theoretical framework and comparing them with the shadow area are provided. The derivation of the NDoF is linked to upper limits on the capacity (spectral efficiency) for a communication channel between an object and the far field, as well as to inverse source problems with measurements surrounding an object.

\appendices
\section{Radiation modes}\label{S:RadiationModes}
Radiation modes for spherical geometries are determined from spherical wave expansions. Homogeneous spherical (ball) region with resistivity $\rhor$ and radius $a$ expressed in spherical waves~\cite{Gustafsson+etal2020} 
\begin{equation}
		\radm_n 
		=\frac{k^2 a^3}{2\rhor}
		\left(
		(\T{R}_{1,l}^{(1)})^2 - \T{R}_{1,l-1}^{(1)}\T{R}_{1,l+1}^{(1)} + \frac{2}{ka}\T{R}_{1,l}^{(1)}\T{R}_{2,l}^{(1)}\delta_{\tau,2}
		\right)
	\label{eq:RadmSphBall}
\end{equation}
and for a spherical shell with surface resistivity $R_{\T{s}}$
\begin{equation}
		\radm_n = \frac{k^2a^2\eta_0}{R_{\T{s}}}(\T{R}_{\tau,l}^{(1)})^2,
	\label{eq:RadmSphShell}
\end{equation}	
where $\T{R}_{\tau,l}$ denotes radial functions~\cite{Hansen1988} evaluated for $ka$.

\section{Directional and polarization averages}\label{S:Averages}
Expressing the solution~\eqref{eq:MaxEff2} in spherical waves with expansion coefficients collected in a column matrix $\M{a}$ using $\M{F}=\M{U}^{\trans}\M{a}/(4\pi)$~\cite{Gustafsson+etal2020} yields 
\begin{equation}
	\max\Aeff = \lambda^2\M{F}^{\herm}\M{R}^{-1}\M{F} 
	=\frac{\lambda^2}{16\pi^2}\M{a}^{\herm}\M{U}\M{R}^{-1}\M{U}^{\trans}\M{a}.
\label{eq:MaxEff2a}
\end{equation}
Diagonalizing~\eqref{eq:MaxEff2a} using radiation modes~\eqref{eq:radmeig} induces a change of variables $\tM{a}=\M{Q}\M{a}$ with a unitary matrix $\M{Q}$ having elements $Q_{nm}$
\begin{equation}
	\max\Aeff
	=\frac{\lambda^2}{16\pi^2}
	\sum_n\frac{\radm_n|\tilde{a}_n|^2}{1+\radm_n}
	=\frac{\lambda^2}{16\pi^2}
	\sum_{n,m}\frac{\radm_n Q^2_{nm}|a_{m}|^2}{1+\radm_n}
\label{eq:MaxEff3}
\end{equation}
with $\sum_m Q^2_{nm}=1$. Taking the average~\eqref{eq:average} of $\max\Aeff$ and
using the plane wave expansion $a_n=4\pi\ju^{\tau-1-l}\UV{e}\cdot\V{Y}_n(\UV{k})$~\cite{Kristensson2016} with spherical harmonics $\V{Y}_n$ and $\UV{e}\cdot\UV{k}=0$ results in
\begin{multline}
	\ave{|a_n|^2}
	=\frac{1}{8\pi^2}\int_{2\pi}\int_{4\pi}|a_n|^2\diff\Omega_{\UV{k}}\diff\Omega_{\UV{e}}\\	=2\int_{4\pi}\V{Y}_\nu(\UV{k})\cdot\int_{2\pi}\UV{e}\UV{e}\diff\Omega_{\UV{e}}\cdot  \V{Y}_\nu(\UV{k})\diff\Omega_{\UV{k}}\\
	=2\pi\int_{4\pi}\V{Y}_\nu(\UV{k})\cdot  \V{Y}_\nu(\UV{k})\diff\Omega_{\UV{k}} = 2\pi,
\label{eq:sphwaveaveint}
\end{multline}
where the middle integral is, without loss of generality, evaluated for an arbitrary fixed direction $\UV{e}=\UV{z}$ as
\begin{equation}
\int_{2\pi}\UV{e}\UV{e}\diff\Omega_{\UV{e}}
=\int_{2\pi}\UV{x}\UV{x}\cos^2\theta+\UV{y}\UV{y}\sin^2\theta\diff\theta
=\pi(\UV{x}\UV{x}+\UV{y}\UV{y})
\label{eq:polpolint}
\end{equation}
and using that $\UV{k}\cdot\V{Y}(\UV{k})=0$ extends~\eqref{eq:polpolint} to an identity dyadic. 
The same relation~\eqref{eq:sphwaveaveint} holds for the orthogonal combination $\ave{|\tilde{a}_n|^2}=\sum Q^2_{mn}\ave{|a_n|^2}=2\pi$ giving the simple identity~\eqref{eq:maxavAff} for the average maximal effective area expressed in radiation modes.


\end{document}